\documentclass[]{pasj01}
\usepackage{times}
\Received{$\langle$reception date$\rangle$}
\Accepted{$\langle$acception date$\rangle$}
\Published{$\langle$publication date$\rangle$}

\SetRunningHead{N. Isobe, \& S. Koyama}
{Electron and magnetic-field energy densities in the west lobe of 3C 236}

\begin{document}
\title{X-ray measurement of electron and magnetic-field energy densities 
       in the west lobe of the giant radio galaxy 3C 236}

\author{
Naoki       \textsc{Isobe}      \altaffilmark{1}, \&
Shoko       \textsc{Koyama}     \altaffilmark{2}
}
\altaffiltext{1}{
        Institute of Space and Astronautical Science (ISAS), 
        Japan Aerospace Exploration Agency (JAXA) \\ 
        3-1-1 Yoshinodai, Chuo-ku, Sagamihara, Kanagawa 252-5210, Japan}
\email{n-isobe@ir.isas.jaxa.jp}
\altaffiltext{2}{
        Max-Planck-Institut f\"{u}r Radioastronomie, 
        Auf dem H\"{u}gel 69, 53121 Bonn, Germany
}


\maketitle
\begin{abstract}
X-ray emission associated with the west lobe of the giant radio galaxy, 3C 236,
was investigated with the Suzaku observatory,
to evaluate the energetics in the lobe.  
After removing contamination from point-like X-ray sources 
detected with Chandra and subtracting the X-ray and non-X-ray backgrounds,
the Suzaku spectrum from the lobe was reproduced 
by a power-low model with a photon index of 
$\Gamma = 2.23_{-0.38-0.12}^{+0.44+0.14}$
where the first and second errors represent  
the statistical and systematic ones, respectively.
Within the errors, 
the X-ray index was consistent with the radio synchrotron one, 
$\Gamma_{\rm R} = 1.74 \pm 0.07$, estimated in the $326$ -- $2695$ MHz range.
This agreement supports that the X-ray emission is attributed 
to the inverse-Compton radiation from the synchrotron electrons 
filling the lobe, 
where the cosmic microwave background photons are up-scattered. 
This result made 3C 236 the largest radio galaxy, 
of which the lobe has ever been probed 
through the inverse-Compton X-ray photons. 
When the photon index was fixed at $\Gamma_{\rm R}$,
the X-ray flux density at 1 keV was measured 
as $S_{\rm X} = 12.3 \pm 2.0 \pm 1.9$ nJy.
A comparison of the X-ray flux to the radio one 
($S_{\rm R} = 1.11 \pm 0.02$ Jy at $608.5$ MHz) yields 
the energy densities of the electrons and magnetic field in the west lobe as  
$u_{\rm e} = 3.9_{-0.7 -0.9}^{+0.6 +1.0} \times 10^{-14} $ ergs cm$^{-3}$
and 
$u_{\rm m} = 0.92_{-0.15 -0.35}^{+0.21 +0.52}\times 10^{-14} $ ergs cm$^{-3}$, 
respectively, 
indicating a mild electron dominance of 
$u_{\rm e}/u_{\rm m} = 4.2_{-1.3 -2.3}^{+1.6 +4.1}$.
The latter corresponds to the magnetic field strength of 
$B = 0.48_{-0.04 -0.10}^{+0.05 +0.12}$ $\mu$G.
These are typical among the lobes of giant radio galaxies. 
A compilation of the $u_{\rm e}$-size relation 
for the inverse-Compton-detected radio galaxies implies that 
the west lobe of 3C 236 is still actively energized by its jet. 
\end{abstract}

\begin{longtable}[t]{llllll}
\caption{Log of observations}
\label{tab:log}
\hline 
Observatory & ObsID     & Date          
            & (R.A., Decl.) \footnotemark[$*$]            & GTI (ks) & Target \\
\hline 
\endfirsthead
\hline 
Observatory & ObsID     & Date          
            & (R.A., Decl.) \footnotemark[$*$]            & GTI (ks) & Target \\
\hline 
\endhead
\hline 
\endfoot
\multicolumn{6}{l}{\footnotemark[$*$] The sky coordinates at the XIS nominal position. } \\
\endlastfoot
Suzaku      & 707005010 & 2012 May 6 -- 8  
            & (\timeform{151D.3326}, \timeform{+35D.0063}) & 78.0 & West Lobe \\
Suzaku      & 707006010 & 2012 May 8 -- 9  
            & (\timeform{151D.1525}, \timeform{+34D.7488}) & 42.4 & XRB       \\
Chandra     & 10246     & 2012 March 10 
            & ---                                         & 29.4 & West Lobe \\
\hline       
\end{longtable}

\section{Introduction}  
\label{sec:intro}
Radio sources with a total projected size of $D \gtrsim 1$ Mpc 
are categorized into giant radio galaxies. 
They constitute one of the largest classes 
of astrophysical objects in the Universe. 
With a typical spectral age $T_{\rm age}$ 
being higher than a few $10$ Myr \citep{GRG_WENSS},
they are recognized as relatively old systems. 
Therefore, the giant radio galaxies are utilized 
to investigate the activity of jets and associated lobes 
in evolved radio sources, which has been yet to be explored. 

Lobes of radio galaxies accumulate 
an enormous amount of relativistic plasma, 
comprised of electrons and magnetic field 
(and possibly heavy particles including protons,
although the present paper does not deal with them). 
The plasma energy in the lobes 
is sourced from the bulk kinetic energy of the jets 
through their terminal hot spots. 
This makes the lobes one of the most valuable 
probes for the past jet activity.
It is able to estimate roughly 
the plasma energy in the lobes by their synchrotron radio radiation.  
However, the radio information alone is not useful to disentangle 
the electron and magnetic-field energies, 
and thus, the minimum energy condition \citep{Bme} was widely assumed,
without any concrete physical justification.  

Another valuable information to study the energetics 
should be provided by inverse Compton (IC) X-ray emission 
from the synchrotron-emitting electrons in the lobes,
where the cosmic microwave background (CMB) photons are boosted up
\citep{CMB_IC}.
The IC X-ray spectrum, in comparison with the synchrotron radio one, 
enables an independent estimate of the electron and magnetic energies. 
In the last two decades
since its discovery from the lobes of Fornax A \citep{ForA_ROSAT,ForA_ASCA},
the IC X-ray radiation 
has been a standard tool to diagnose the lobe energetics 
(e.g., 
\cite{CenB,3C452,3C98,lobes_Croston,ForA,PicA_XMM,ForA_Suzaku,NGC6251,CenA}). 
The technique was strengthen 
by the IC $\gamma$-ray detection with the Fermi observatory
from the outer lobes of Centaurus A  \citep{CenA_Fermi}. 
As a result, the electron energy is found to surpass the magnetic one 
in the lobes, typically by a factor of $1$ -- $100$. 

Until recently, 
IC X-ray examination on lobes of giant radio galaxies
has been hampered by mainly their faintness. 
For diffuse X-ray sources with a low surface brightness, 
the X-ray Imaging Spectrometer (XIS; \cite{XIS}) 
onboard the Suzaku observatory \citep{Suzaku} exhibits a great advantage, 
owing to its moderate field of view (a $\sim$\timeform{18'} square) 
and to its rather low and stable instrumental background \citep{xisnxbgen}. 
Actually, 
a systematic study with the XIS has been just conducted 
for the lobes of several giant radio galaxies \citep{3C326,3C35,DA240}.
This research arrived at a significant detection of IC X-ray photons 
from these lobes, and pioneeringly hinted that the current jet activities are 
significantly declined in the giant radio galaxies,
compared with those with a moderate size (e.g., $D \lesssim 500$ kpc). 
This motivates us to enlarge urgently 
the sample of IC-detected giant radio galaxies. 

The giant radio galaxy, 3C 236, is located 
at the redshift of $z = 0.1005 \pm 0.0005$ \citep{3C236_redshift}.
Assuming the cosmology with $H_{\rm 0} = 71$ km s$^{-1}$ Mpc$^{-1}$, 
$\Omega_{\rm m} = 0.27$, and  $\Omega_{\lambda} = 0.73$,
the redshift gives 
an angle-to-size conversion ratio of $109.8{\rm ~kpc}/$\timeform{1'}. 
It is optically classified as a low excitation radio galaxy, 
with weak emission lines \citep{RG_178MHz}. 
With a classical Fanaroff-Riley II radio morphology \citep{RG_178MHz},
the radio source has an angular extent of $\sim$\timeform{40'}
\citep{3C236_largest,3C236_largest2},
which corresponds to a projected linear size of $D \sim 4.4$ Mpc. 
Therefore, this had been known as the largest radio source all over the sky,
until the giant radio galaxy J1420-0545 with a size of $D\sim 4.7$ Mpc 
was discovered \citep{J1420}. 
The age of the source is estimated as $T_{\rm age} = 98\pm3$ Myr 
by the synchrotron aging technique \citep{GRG_WENSS},
indicating that it is really an elderly radio galaxy. 
These make 3C 236 as an ideal target for the IC X-ray study with Suzaku. 
Actually, a faint X-ray emission associated with its west lobe 
has been concretely detected in the Suzaku observation,
as is reported in the present paper. 

\section{Observation and Data Reduction} 
\label{sec:obs}
\subsection{Suzaku Observation}
\label{sec:obs_suzaku}
The log of the Suzaku observations for the giant radio galaxy, 3C 236,
is given in table \ref{tab:log}.
We focused on the west lobe, 
since radio images of 3C 236 (e.g., \cite{3C236_radio_structure,GRG_image})
indicate that 
diffuse radio emission is more pronounced in the west lobe
while a significant dominance of double bright hot spots is 
recognized in the east one.
The central region of the west lobe was placed 
at the nominal position of the X-Ray Telescope (XRT; \cite{XRT}) for the XIS.
We fixed the satellite roll angle at $300$ deg,
in order to avoid contamination onto the west lobe 
from calibration sources at the specific corners of the XIS field of view,
and distortion by the anomalous columns of the 
XIS 0\footnote{See \S 7.1.2 of the Suzaku Technical Description;\\
{\tt http://www.astro.isas.jaxa.jp/suzaku/doc/suzaku\_td/}.}. 
Unfortunately, in this configuration,
the 3C 236 nucleus is located slightly outside the XIS field of view. 
We observed another sky field on the southwest of the west lobe field,
which is known to be relatively free from 
contaminating bright X-ray point sources (e.g., \cite{1RXS}),
and evaluated the X-Ray Background (XRB) spectrum. 
This XRB field has not yet been observed with ASCA, Chandra, and XMM-Newton.

The present paper deals with the XIS results only, 
because no significant X-ray signals were detected 
with the Hard X-ray Detector \citep{HXD} from both sky field.  
The standard software package, HEASoft 6.16, 
was adopted for data reduction and analysis. 
We referred the calibration database (CALDB) 
as of 2014 July 1 and 2011 June 30 for the XIS and XRT, respectively. 
All the XIS events were reprocessed with the Suzaku tool, {\tt aepipeline}. 
We filtered the data by the following standard criteria;
the satellite is outside the South Atlantic Anomaly (SAA), 
the time after an SAA passage is larger than $436$ s, 
the geomagnetic cut-off rigidity is higher than 6 GV, 
the source elevation is higher than \timeform{20\circ} and \timeform{5\circ}
above the rim of day and night Earth, respectively,
and the XIS data are unaffected by telemetry saturation.
The filtering procedures yielded 78.0 ks and 42.4 ks of good exposures 
for the west-lobe and XRB fields, respectively. 

In the scientific analysis below, 
we picked up those events with an XIS grade of 0, 2, 3, 4, or 6.
In order to mitigate the artificial increase 
in the Non-X-ray Background (NXB) level related to the charge injection,
those pixels flagged as SCI\_2ND\_TRAILING\_ROW were masked out
for the backside-illuminated (BI) CCD chip 
(XIS 1)\footnote{See 
``Amount of Charge Injection for XIS 1 and NXB Increase with CI=6 keV''; \\
{\tt http://www.astro.isas.jaxa.jp/suzaku/analysis/xis/xis1\_ci\_6\_nxb/}}.

\begin{figure*}[t]
\centerline{
  \FigureFile(80mm,80mm){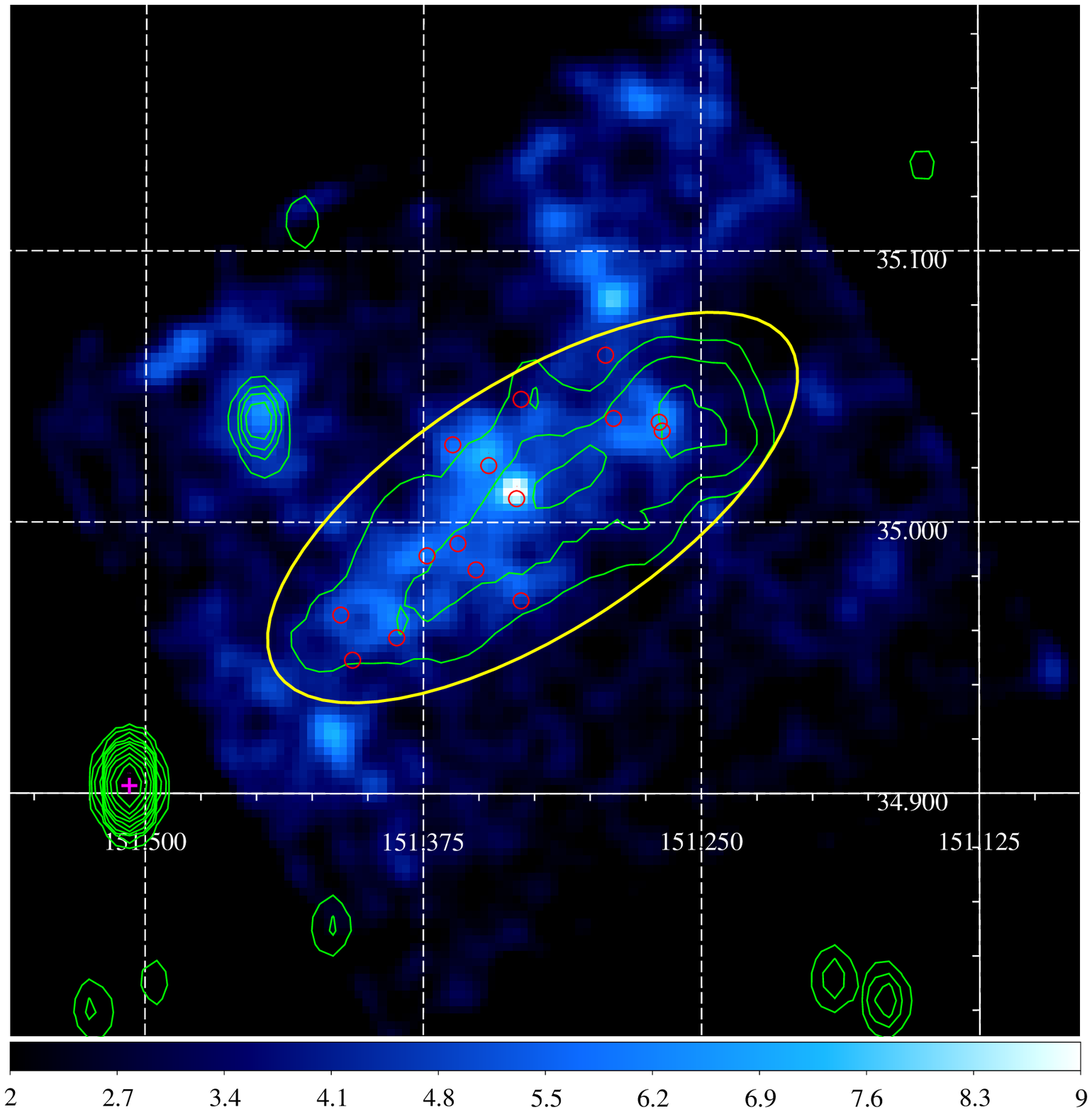}
  \hspace{0.3cm}
  \FigureFile(80mm,80mm){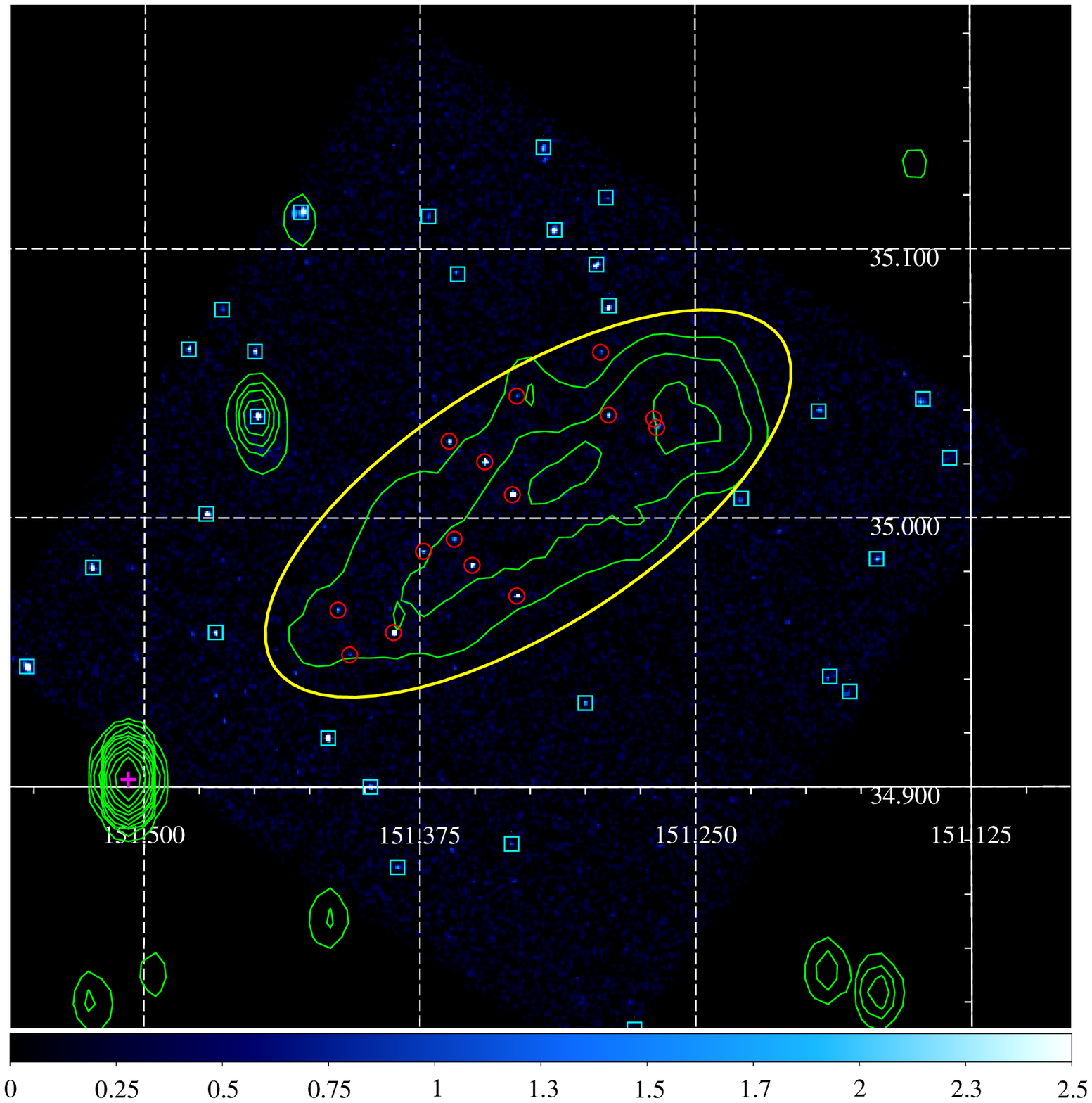}
}
\caption{
(left) 
Background-inclusive Suzaku XIS image of the 3C 236 west lobe 
in the 0.5 -- 5.5 keV range,
on which the 608.5 MHz radio contours 
\citep{GRG_image} are superposed.
The XIS image was smoothed with a two-dimensional Gaussian kernel 
of a \timeform{30"} radius, while it was uncorrected for exposure. 
The scale bar indicates the signal counts 
within a \timeform{10"} $\times$ \timeform{10"} bin. 
The XIS spectrum of the west lobe was accumulated from the solid ellipse.
The circles point the contaminating Chandra ACIS sources,
taken into account in the XIS spectral analysis 
(see tables \ref{tab:src} and \ref{tab:spec_SRC}).
The nucleus of 3C 236, shown with the cross, 
is located outside the XIS field of view.  
(right) 
Background-inclusive Chandra ACIS image of the similar sky field 
in $0.5$ -- $7$ keV, smoothed with a two-dimensional Gaussian 
of a \timeform{4"} radius. 
The scale bar displays the signal counts 
within a \timeform{2"} $\times$ \timeform{2"} bin. 
The X-ray sources detected with the ACIS 
inside the XIS spectral integration region (the ellipse) 
are overplotted with the circles, 
while those outside the region are indicated by the boxes.
} 
\label{fig:img_WL}
\end{figure*}

\subsection{Archival Chandra Data}      
\label{sec:obs_chandra}
Thanks to its sub-arcsecond angular resolution,
Chandra is very useful to evaluate 
the X-ray flux of possible contaminating faint point-like sources 
located within the west lobe, which are unresolved with the Suzaku XIS.  
3C 236 has ever been observed by the Chandra ACIS-I array 
with three exposures. 
Among them, we selected the observation 
aiming at its west lobe (ObsID = 10246; see table \ref{tab:log}). 

The ACIS data were reduced with the software package CIAO 4.6. 
We reprocessed the data and created a new level 2 event file 
with the tool {\tt chandra\_repro}, by referring to CALDB 4.6.2.
Because the NXB count rate integrated over the whole ACIS-I array 
was found to be fairly stable throughout the observation,
we performed no additional data screening to the new level 2 event file.
Correspondingly, we obtained 29.4 ks of good exposure on the west lobe.
When the science products were derived, 
the grade selection criterion similar to that for the Suzaku XIS
(i.e., 0, 2, 3, 4, or 6) was applied.

\begin{table}[t]
\caption{Chandra X-ray sources detected inside the ellipse,
representing the west lobe.}
\label{tab:src}
\begin{center}
\begin{tabular}{llr}
\hline 
(R.A., Decl.)                  & $\Delta\theta$ \footnotemark[$*$] 
                               & $t$ \footnotemark[$\dagger$] \\  
\hline 
(\timeform{151D.29288}, \timeform{+35D.06163}) & $ 1.2 $ & $ 3.2 $   \\ 
(\timeform{151D.33095}, \timeform{+35D.04535}) & $ 0.7 $ & $ 4.5 $   \\ 
(\timeform{151D.28930}, \timeform{+35D.03829}) & $ 0.3 $ & $ 6.4 $   \\ 
(\timeform{151D.26875}, \timeform{+35D.03690}) & $ 0.9 $ & $ 4.3 $   \\ 
(\timeform{151D.26740}, \timeform{+35D.03364}) & $ 0.8 $ & $ 8.0 $   \\ 
(\timeform{151D.36171}, \timeform{+35D.02858}) & $ 0.1 $ & $ 10.7 $  \\ 
(\timeform{151D.34551}, \timeform{+35D.02092}) & $ 0.1 $ & $ 30.7 $  \\ 
(\timeform{151D.33296}, \timeform{+35D.00870}) & $ 0.1 $ & $ 18.9 $  \\ 
(\timeform{151D.35938}, \timeform{+34D.99214}) & $ 0.2 $ & $ 4.9 $   \\ 
(\timeform{151D.37328}, \timeform{+34D.98764}) & $ 0.2 $ & $ 8.6 $   \\ 
(\timeform{151D.35123}, \timeform{+34D.98246}) & $ 0.1 $ & $ 10.1 $  \\ 
(\timeform{151D.33097}, \timeform{+34D.97106}) & $ 0.1 $ & $ 22.4 $  \\ 
(\timeform{151D.41207}, \timeform{+34D.96578}) & $ 0.3 $ & $ 6.1 $   \\ 
(\timeform{151D.38690}, \timeform{+34D.95742}) & $ 0.1 $ & $ 16.4 $  \\ 
(\timeform{151D.40676}, \timeform{+34D.94915}) & $ 0.6 $ & $ 3.6 $   \\ 
\hline 
\multicolumn{3}{@{}l@{}}{\hbox to 0pt{\parbox{80mm}{\footnotesize
\vspace{0.2cm}
\par\noindent\footnotemark[$*$] The position error in arcsec 
\par\noindent\footnotemark[$\dagger$] The detection significance 
}\hss}}
\end{tabular}
\end{center}
\end{table}

\section{Image Analysis}  
\label{sec:image}
The left panel of figure \ref{fig:img_WL} 
displays the XIS image of the 3C 236 west lobe.
Here, we extracted the XIS events in the 0.5 -- 5.5 keV range, 
where the observed data are free from photons 
from the calibration sources at the field-of-view corners. 
The data from all the CCD chips (XIS 0, 1 and 3) were summed up. 
Neither the XRB nor NXB was subtracted.
The image was smoothed with a two-dimensional Gaussian kernel 
of \timeform{30''} radius. 
The radio image of 3C 236 at 608.5 MHz 
\citep{GRG_image}\footnote{The electric version is available 
from the NASA/IPAC Extragalactic Database (NED),
operated by the Jet Propulsion Laboratory, California Institute of Technology, 
under contract with the National Aeronautics and Space Administration.}
was superposed with contours.
The figure \ref{fig:img_WL} left panel  
suggests faint X-ray emission associated with the west lobe, 
although it is certainly contaminated by several unresolved X-ray sources 
behind the lobe. 

The Chandra ACIS-I image of the same sky field in $0.5$ -- $7$ keV 
is presented in the right panel of figure \ref{fig:img_WL}.
The image clearly reveals numbers of point-like X-ray sources.  
We performed a source detection procedure 
by the CIAO tool {\tt wave\_detect} with a standard parameter set. 
The exposure and point-spread-function maps 
over the ACIS field of view was correctly taken into account 
for the source detection.
In total, $44$ X-ray sources were detected 
with a significance of $t>3\sigma$ within the ACIS field of view. 
These sources are plotted with the circles or boxes in figure \ref{fig:img_WL}. 

Utilizing the 608.5 MHz radio map, we evaluated the size of the west lobe. 
The standard deviation, $\sigma_{\rm r}$, 
of the radio intensity in the source free pixels 
was regarded as the noise level of the radio map. 
The emission region was defined as those pixels  
with a radio surface brightness of $\ge 3 \sigma_{\rm r}$.
We successfully approximated 
the envelope of the emission region of the west lobe 
by an ellipse, which is shown in figure \ref{fig:img_WL} with the solid line. 
After decomvolving the radio beam size,
the ellipse has a major and minor radius of 
\timeform{6'.74} and \timeform{2'.78}, respectively,  
corresponding to the projected size of $740$ kpc and $305$ kpc 
at the redshift of 3C 236.
The X-ray spectrum from the west lobe is accumulated in this ellipse
(see \S\ref{sec:WestLobe}). 
On the region, the $15$ Chandra sources,
plotted with the circles in figure \ref{fig:img_WL},
were detected. 
These sources are listed in table \ref{tab:src}. 

\begin{figure}[h]
\centerline{
\FigureFile(80mm,80mm){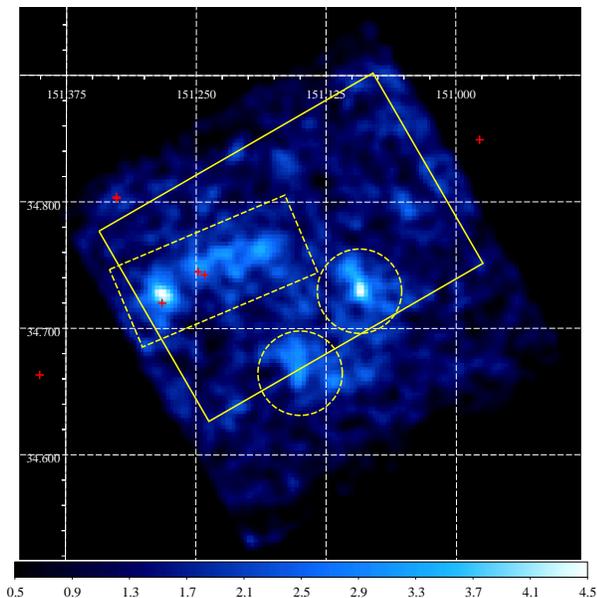}}
\caption{Suzaku XIS image of the XRB region in 0.5 -- 5.5 keV,
smoothed with a two-dimensional Gaussian function of a \timeform{30''} radius.
The integration region for the XRB specrum is indicated by the solid rectangle,
from which the dashed rectangle and two circles were removed. 
The X-ray sources, taken from the Second ROSAT Source Catalog 
of Pointed Observations with the Position Sensitive Proportional Counter,
are displayed with the crosses. 
}
\label{fig:img_XRB}
\end{figure}

\begin{table}[h]
\caption{Best-fit spectral parameters for the XRB}
\label{tab:spec_XRB}
\begin{center}
\begin{tabular}{ll}
\hline 
Parameter             &  Value\\
\hline 
$N_{\rm H}$ (cm$^{-1}$) & $9.3 \times 10^{19}$ \footnotemark[$\S$] \\
$\Gamma$              & $1.41$\footnotemark[$\#$] \\
$kT_{\rm 1}$           & $0.204$ \footnotemark[$**$] \\
$kT_{\rm 2}$           & $0.074$ \footnotemark[$**$] \\
$f_{\rm tot}$  (erg s$^{-1}$ cm$^{-2}$ str$^{-1}$) \footnotemark[$*$] 
                      & $(6.1 \pm 0.2)\times 10^{-8}$ \\
$f_{\rm PL}$ (erg s$^{-1}$ cm$^{-2}$ str$^{-1}$) \footnotemark[$\dagger$] 
                      & $4.6 \times 10^{-8}$\\
$f_{\rm th}$ (erg s$^{-1}$ cm$^{-2}$ str$^{-1}$) \footnotemark[$\ddagger$] 
                      & $2.5 \times 10^{-8}$\\
$\chi^2/{\rm dof}$    & $95.8/79$\\
\hline        
\multicolumn{2}{@{}l@{}}{\hbox to 0pt{\parbox{68mm}{\footnotesize
\vspace{0.3cm}
\par\noindent\footnotemark[$*$] 
    The total absorption-inclusive surface brightness in $0.5$ -- $5$ keV.
\par\noindent\footnotemark[$\dagger$] 
    The absorption-inclusive surface brightness 
    of the PL component in $2$ -- $10$ keV.  
\par\noindent\footnotemark[$\ddagger$] 
    The absorption-inclusive surface brightness in $0.5$ -- $2$ keV,
    summed over the 2 MEKAL components. 
\par\noindent\footnotemark[$\S$] Fixed at the Galactic value \citep{NH}.
\par\noindent\footnotemark[$\#$] Taken from \citet{XRB_ASCA}.
\par\noindent\footnotemark[$**$] Taken from \citet{XRB_XMM}.
}\hss}}\end{tabular}
\end{center}
\end{table}

\section{Spectral Analysis}
\label{sec:spec}
\subsection{X-ray Background}
\label{sec:XRB}
It is inevitably required to precisely subtract the NXB and XRB levels,
for the spectral analysis of extended X-ray emission 
with a low surface brightness. 
It is reported that the NXB level of the XIS is predictable 
with an accuracy of $\sim 3$\% for a typical exposure of $50$ ks
by utilizing the {\tt xisnxbgen} tool \citep{xisnxbgen}.
We make use of the XIS data accumulated from the XRB field 
(see table \ref{tab:log})
to model the XRB spectrum in the direction to 3C 236. 

\begin{figure}[t]
\centerline{\FigureFile(80mm,80mm){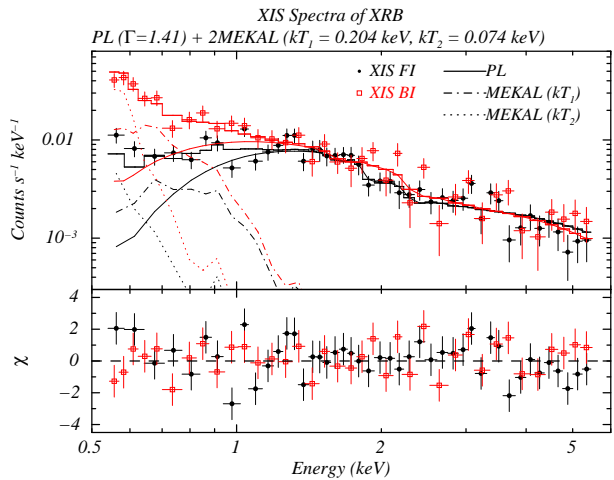}}
\vspace{0.2cm}
\caption{NXB-subtracted Suzaku XIS spectreum of the XRB. 
The best-fit model, comprised of a PL component ($\Gamma = 1.41$; solid line)
and two MEKAL ones ($kT_1 = 0.204$ keV and $kT_2 = 0.074$ keV; 
the dashed-dotted and dotted lines respectively), 
is indicated by the histograms.}
\label{fig:spec_XRB}
\end{figure}

The $0.5$ -- $5.5$ keV XIS image of the XRB field is plotted 
in figure \ref{fig:img_XRB}. 
We extracted the XRB spectrum within the solid rectangle in the image.
The region is selected so as not to be affected 
by the anomalous columns of the XIS 0.
In addition to the previously known X-ray sources 
(the crosses in figure \ref{fig:img_XRB}),
listed in the Second ROSAT Source Catalog of Pointed Observations
with the Position Sensitive Proportional Counter\footnote{
Taken from {\tt http://www.xray.mpe.mpg.de/cgi-bin/rosat/src-browser}},
a few point sources and faint extended X-ray emission 
were recognized within the regions.  
Then, the dashed rectangle and circles in the image were discarded. 
Figure \ref{fig:spec_XRB} shows the XRB spectrum derived from the region,
after the NXB spectrum generated by {\tt xisnxbgen} was subtracted.  
We co-added the data from the two front-illuminated (FI) 
CCD chips (XIS 0 and 3). 
The XIS response matrix functions (rmfs) 
of this region were created by {\tt xisrmfgen}.
Assuming a diffuse source with a radius of \timeform{20'},
which has a flat spatial distribution, 
we simulated the auxiliary response files (arfs) by {\tt xissimarfgen} 
\citep{xissimarfgen}. 

The $0.2$ -- $10$ keV XRB spectrum is widely known to be decomposed into 
a power-law (PL) component and a two-temperature thermal plasma emission 
\citep{XRB_ASCA,XRB_XMM}. 
The PL component is reported to exhibit 
an average surface brightness of 
$ f_{\rm PL} = (6.4 \pm 0.6) \times 10^{-8}$ ergs cm$^{-2}$ s$^{-1}$ str$^{-1}$
in $2$ -- $10$ keV, 
with a rather small spatial fluctuation of $\sim 7$\%.
Its photon index is accurately determined 
as $\Gamma = 1.41$ \citep{XRB_ASCA}. 
This component is believed to be dominated by unresolved faint sources,
including distant active galactic nuclei.
By adopting the MEKAL code \citep{MEKAL},
the temperatures of the thermal plasma emission were determined 
as $kT_{\rm 1} = 0.204$ keV and $kT_{\rm 2} = 0.074$ keV \citep{XRB_XMM}. 
The thermal 
components are
thought to be of Galactic and local origin,
and its surface brightness is reported to be highly variable 
from field to field.

We fitted the XIS spectrum from the XRB region
with a composite model consisting of the PL component ($\Gamma = 1.41$)
and the two MEKAL ones ($kT_{\rm 1} = 0.204$ keV and $kT_{\rm 2} = 0.074$ keV). 
All the components were subjected 
to the Galactic absorption toward 3C 236
with a hydrogen column density of $N_{\rm H} = 9.3 \times 10^{19}$ cm$^{-2}$
\citep{NH}. 
The model was found to be reasonable ($\chi^2/{\rm dof} = 95.8/79$), 
yielding the best-fit parameters tabulated in table \ref{tab:spec_XRB}. 
The measured $2$ -- $10$ keV surface brightness of the PL component,
$f_{\rm PL } = 4.6 \times 10^{-8}$ ergs cm$^{-2}$ s$^{-1}$ str$^{-1}$, 
falls near the lowest end of the XRB surface brightness distribution 
\citep{XRB_ASCA}.
Therefore, we have judged that the best-fit XRB model is safely applied 
to the spectral analysis of the west lobe. 

\subsection{Contaminating point sources}
\label{sec:Src}
The Chandra ACIS signals from each X-ray source, 
detected on the integration region for the XIS spectrum 
from the west lobe (the ellipse in figure \ref{fig:img_WL}),
were integrated within a circle of a $10$ pixel (\timeform{4.9"}) radius 
centered on the source position (see table \ref{tab:src}).
The NXB plus XRB events were accumulated from 
a concentric annuls with a radius of $15$ -- $30$ pixel 
(\timeform{7.4''}--\timeform{14.8"}).
The spectral files, together with the rmf and arf ones, 
were created by the CIAO script {\tt specextract}.
The energy-dependent aperture correction was applied to the arfs. 

\begin{figure}[h]
\centerline{\FigureFile(80mm,80mm){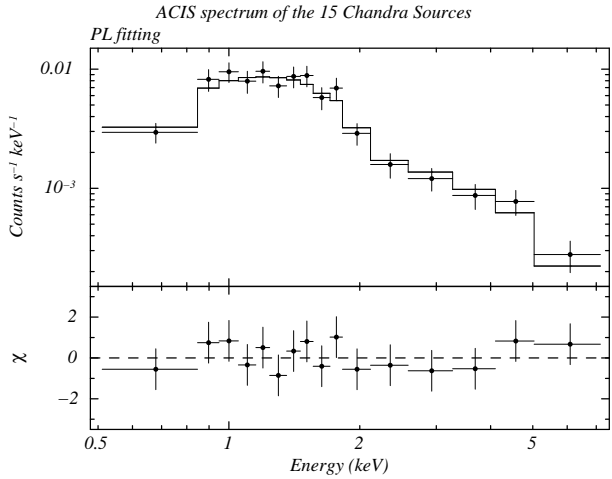}}
\vspace{0.2cm}
\caption{Sum of the Chandra ACIS spectra from the 15 sources 
detected on the west lobe (table \ref{tab:src}).
The histogram indicates the best-fit PL model. }
\label{fig:spec_SRC}
\end{figure}

\begin{table}[t]
\caption{Best-fit parameters for the summed spectrum 
         of the 15 contaminating sources on the west lobe.}
\label{tab:spec_SRC}
\begin{center}
\begin{tabular}{ll}
\hline 
Parameter             &  Value\\
\hline 
$N_{\rm H}$ (cm$^{-1}$) & $1.3_{-1.0}^{+1.2} \times 10^{21}$ \\
$\Gamma$              & $2.0 \pm 0.3 $\\
$F_{\rm src}$ (erg s$^{-1}$ cm$^{-2}$)   \footnotemark[$*$]        
                      & $(1.2\pm0.1)\times 10^{-13}$\\
$\chi^2/{\rm dof}$    & $ 6.9/13 $\\
\hline        
\multicolumn{2}{@{}l@{}}{\hbox to 0pt{\parbox{70mm}{\footnotesize
\par\noindent\footnotemark[$*$] The $0.5$-- $5$ keV absorption-inclusive flux 
}\hss}}\end{tabular}
\end{center}
\end{table}

Due to low signal statistics for the individual sources, 
we presents the ACIS spectrum summed over the 15 sources
in figure \ref{fig:spec_SRC}. 
The rmf was simply averaged over these sources,
after it was multiplied by the corresponding arf. 
The spectrum was successfully approximated by a PL model 
modified by a free absorption ($\chi^2/{\rm dof} = 6.9/13$).
The best-fit parameters are summarized in table \ref{tab:spec_SRC}. 
The summed absorption-inclusive X-ray flux from the sources 
was measured as 
$F_{\rm src} = (1.2\pm0.1) \times 10^{-13}$ erg s$^{-1}$ cm$^{-2}$
in $0.5$ -- $5$ keV. 
The photon index, $\Gamma = 2.0 \pm 0.3 $,
was found to agree with a typical value of nearby active galaxies 
in the similar energy range \citep{AGN_Suzaku}.

\begin{longtable}[t]{lllllll}
\caption{Signal statistics from the west lobe region.}
\label{table:stat:WL}
\hline 
           & \multicolumn{3}{l}{FI signal ($10^{-2}$ cts s$^{-1}$)\footnotemark[$*$] } 
           & \multicolumn{3}{l}{BI signal ($10^{-2}$ cts s$^{-1}$)\footnotemark[$*$] } \\
\hline
\endfirsthead
\hline 
           & \multicolumn{3}{l}{FI signal ($10^{-2}$ cts s$^{-1}$)\footnotemark[$*$] } 
           & \multicolumn{3}{l}{BI signal ($10^{-2}$ cts s$^{-1}$)\footnotemark[$*$] } \\
\hline
\endhead
\hline 
\endfoot
\multicolumn{7}{l}{\footnotemark[$*$]        Count rate per one CCD chip evaluated in $0.6$ - $5$ keV.} \\
\multicolumn{7}{l}{\footnotemark[$\dagger$]  Statistical error.} \\
\multicolumn{7}{l}{\footnotemark[$\ddagger$] Systematic error.}  \\
\multicolumn{7}{l}{\footnotemark[$\S$]       Contamination from the 15 Chandra srouces detected on the west lobe region.}  \\ 
\multicolumn{7}{l}{\footnotemark[$\#$]       Excess signals, after the NXB, XRB and source contamination were subtracted.} \\
\endlastfoot
Data       & $2.44$  & $0.04$   & ---
           & $3.65$  & $0.07$   & --- \\
NXB        & $0.50$  & $0.005$  & $0.015$ 
           & $0.98$  & $0.02$   & $0.03$ \\
Signal     & $1.94$  & $0.04$   &  $0.015$  
           & $2.67$  & $0.07$   &  $0.03$  \\ 
\hline     
XRB        & $1.01$  &  ---     & $0.03$ 
           & $1.44$  &  ---     & $0.04$ \\
Sources\footnotemark[$\S$]   
           & $0.58$  &  ---     & $0.03$
           & $0.73$  &  ---     & $0.04$ \\
\hline        
Excess\footnotemark[$\#$]     
           & $0.35$  & $0.04$   & $0.05$ 
           & $0.50$  & $0.07$   & $0.07$ \\
\hline        
\end{longtable}

\subsection{West Lobe}
\label{sec:WestLobe}
\subsubsection{Significant excess signals} 
\label{sec:WestLobe_stat}
\begin{figure*}[t]
\centerline{
\FigureFile(75mm,75mm){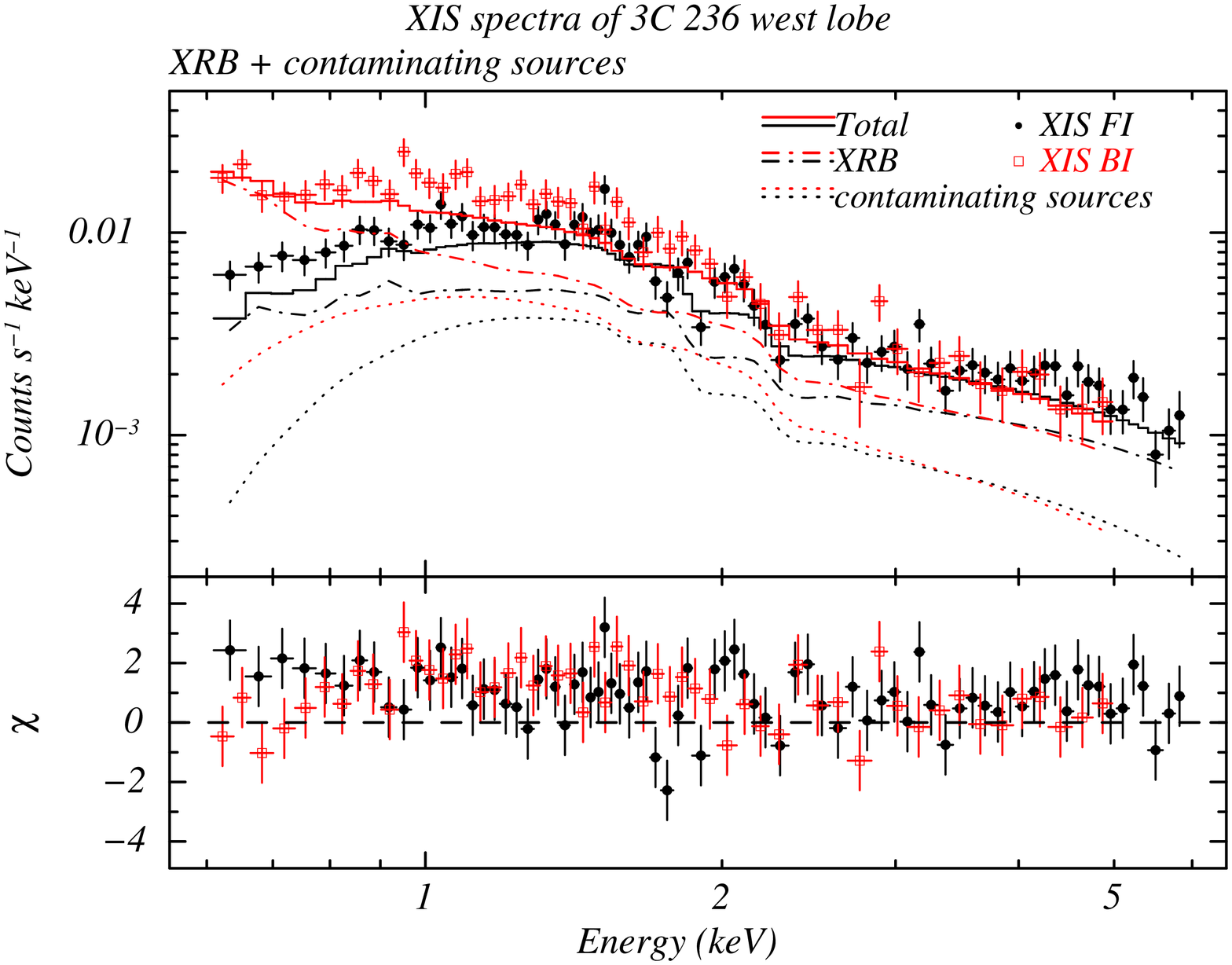}
\hspace{0.5cm}
\FigureFile(75mm,75mm){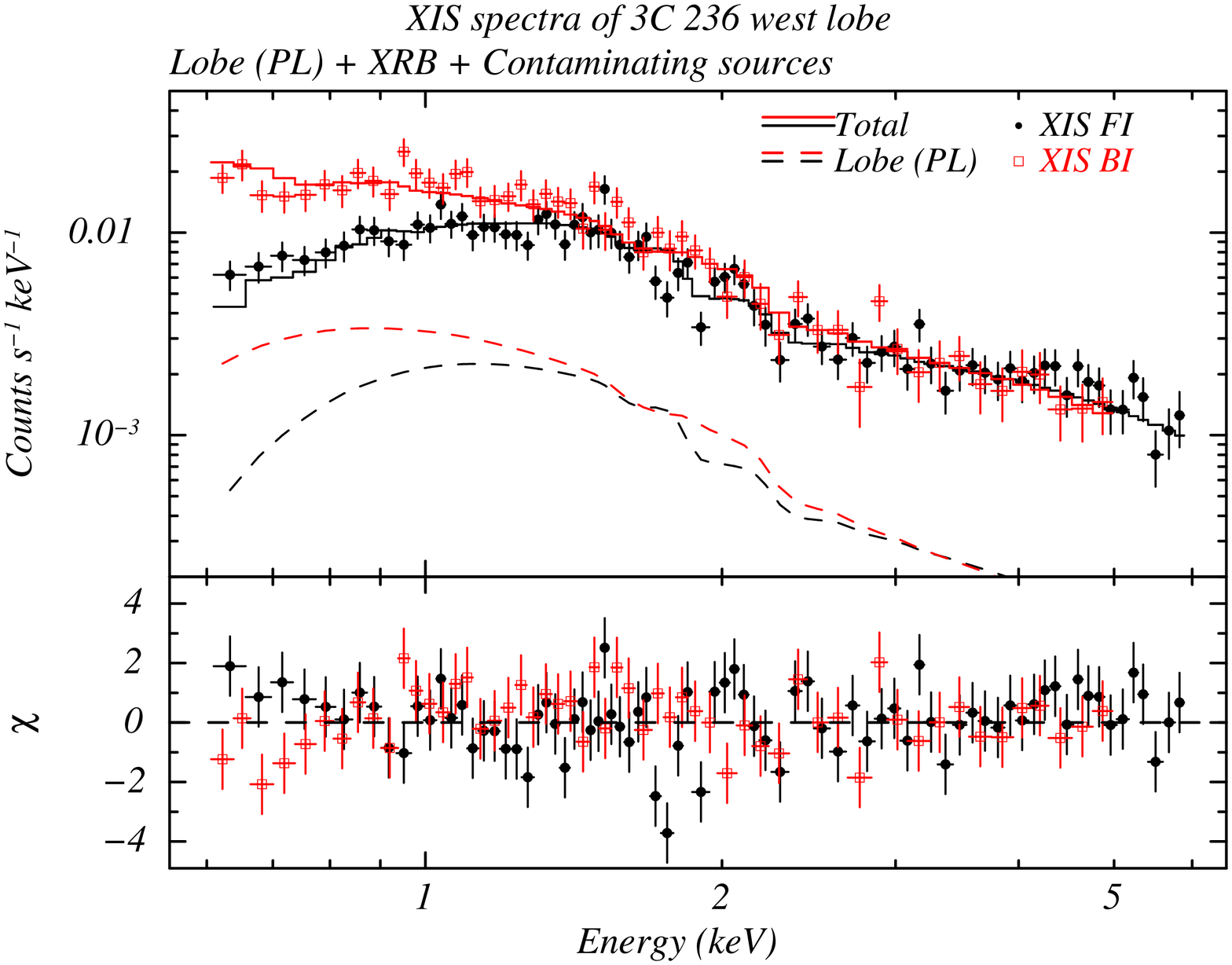}}
\vspace{0.2cm}
\caption{
NXB-subtracted XIS spectra of the 3C 236 west lobe.
In the left panel, the XIS data are compared with the spectral model 
(the solid histogram) consisting of the XRB (the dased-dotted lines) 
and the contamination from the Chandra X-ray sources 
detected on the west lobe (the dotted lines). 
The residual spectrum is plotted in the bottom layer. 
In the right panel, the excess over the XRB plus conamination source model
is fitted by an additional PL component (the dashed lines),
representing the X-ray emission from the west lobe.
}
\label{fig:spec_WL}
\end{figure*}

Figure \ref{fig:spec_WL} shows the XIS spectra of the west lobe 
integrated within the ellipse in figure \ref{fig:img_WL},
after the NXB events simulated by the {\tt xisnxbgen} tool were subtracted.
The data below 0.6 keV were discarded,
since we noticed considerable inconsistency between the XIS FI and BI data 
in the spectral fitting over this energy range,
probably due to the calibration uncertainty.
The signal statistics from the region in the $0.6$ -- $5.0$ keV range
are summarized in table \ref{table:stat:WL}. 
We adopted the typical NXB systematic uncertainty of $3$\% \citep{xisnxbgen}.
Significant X-ray signals exceeding the NXB were detected, 
with an FI and BI count rate per one CCD chip 
of 
$(1.94 \pm 0.04 \pm 0.015) \times 10^{-2}$ cts s$^{-1}$
and 
$(2.67 \pm 0.07 \pm 0.03) \times 10^{-2}$ cts s$^{-1}$,
respectively.
Here, the first and second error represents 
the statistical and systematic one, respectively. 

The NXB-subtracted XIS spectra of the west lobe 
plotted in figure \ref{fig:spec_WL} 
inevitably include the XRB signals and contamination from the point sources,
in addition to the signals from the west lobe itself. 
In the similar manner to that adopted in \S \ref{sec:XRB},
the rmf and arf of the west lobe region to the XRB were generated.
The dash-dotted lines in the left panel of figure \ref{fig:spec_WL}
indicate the XRB model spectrum (table \ref{tab:spec_XRB}), 
which is convolved with the XIS rmf and arf. 
The FI and BI count rates of the XRB within the west lobe region 
in $0.6$ -- $5.0$ keV were estimated as 
$(1.01 \pm 0.03) \times 10^{-2}$ cts s$^{-1}$
and 
$(1.44 \pm 0.04) \times 10^{-2}$ cts s$^{-1}$,
respectively.

The rmfs and arfs of the west lobe region 
to the $15$ Chandra X-ray sources
listed in table \ref{tab:src} were individually calculated,
and they were simply averaged over all the sources. 
The best-fit PL model to the sum of the Chandra spectra from these sources 
(see figure \ref{fig:spec_SRC} and table \ref{tab:spec_SRC})
is drawn with the dotted lines in the figure \ref{fig:spec_WL} left panel,
after it is convolved with the averaged rmf and arf. 
The contamination from these sources 
onto the $0.6$ -- $5.0$ keV FI and BI count rate of the west lobe 
were estimated as $ (0.58 \pm 0.03) \times 10^{-2}$ cts s$^{-1}$
and $ (0.73 \pm 0.04) \times 10^{-2}$ cts s$^{-1}$,
respectively. 
We neglected the Chandra sources 
detected outside the signal integration region,
since their contribution is inferred as 
at most $9$\% of the sum signals 
from the $15$ sources. 

The solid lines in the left panel of figure \ref{fig:spec_WL} 
present the sum spectrum of the XRB and source contamination. 
The residuals plotted in the bottom clearly visualizes 
excess signals over these two components 
($\chi^2/{\rm dof} = 249.1/134$).
The $0.6$ -- $5.0$ keV excess count rate is estimated 
as $(0.35 \pm 0.04 \pm 0.05) \times 10^{-2}$ cts s$^{-1}$ 
and $(0.50 \pm 0.07 \pm 0.07) \times 10^{-2}$ cts s$^{-1}$ 
with the XIS FI and BI, respectively.
Here and hereafter, 
the errors of the best-fit spectral models 
to the XRB and to the contaminating sources 
were propagated to the systematic error (i.e., the second one). 
Thus, 
the statistical significance of the excess with the FI and BI
was evaluated as $8.7 \sigma$ and $7.4 \sigma$, 
respectively.
Even if the systematic errors are taken into account, 
the excess has remained meaningful.

\subsubsection{Spectral modeling} 
\label{sec:WestLobe_fitting}
\begin{figure}[h]
\centerline{\FigureFile(75mm,75mm){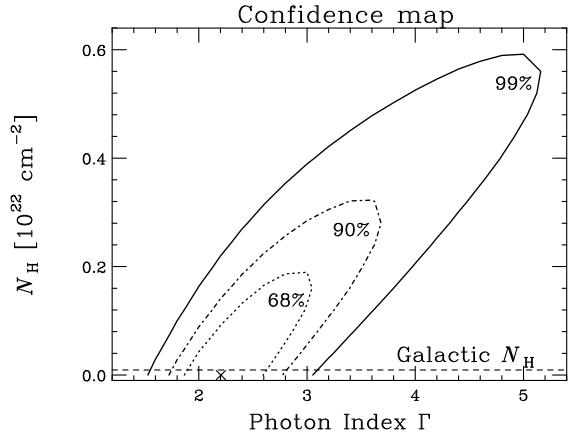}}
\vspace{0.2cm}
\caption{Confidence contours on the $\Gamma$ -- $N_{\rm H}$ plane 
for the additional PL component 
fitted to the excess XIS spectrum of the west lobe.
Three confidence levels are indicated at 68\%, 90\% and 99\%,
with the dotted, dash-dotted and solid lines, respectively.
The cross points the best-fit values 
($\Gamma = 2.20$ and $N_{\rm H} = 3.0 \times 10^{16}$ cm$^{-2}$). 
The horizontal dashed line corresponds to the Galactic column density,
$N_{\rm H} = 9.3 \times 10^{19}$ cm$^{-2}$.} 
\label{fig:confidence}
\end{figure}

In order to reproduce the excess spectrum from the west lobe region,
we employed an additional absorbed PL component. 
The arf to this PL component is 
calculated for a diffuse source filling the ellipse in figure \ref{fig:img_WL}
with a uniform surface brightness. 
First, we fitted the XIS spectrum 
with the photon index $\Gamma$ and absorption column density $N_{\rm H}$ 
both left free.
The confidence map in the figure \ref{fig:confidence} visualizes that 
the Galactic absorption toward 3C 236 with the column density 
of $N_{\rm H} = 9.3 \times 10^{19}$ cm$^{-2}$ \citep{NH} is fairly preferable,
although the acceptable range is found to be slightly wide.
\citet{3C236_outflow} reported a massive nuclear outflow of neutral gas 
with a column density of $N_{\rm H} = 5 \times 10^{21}$ cm$^{-2}$,
which could affect the nuclear jet emission.
Such a high column density,
which requires a very soft spectrum ($\Gamma \gtrsim 4$), 
is outside the $98$\% statistical confidence range. 
We think that the nuclear outflow should not impinge upon 
the X-ray emission from the Mpc-scale lobe,
since it is reported to exhibit a spatial extent of only $\sim 1$ kpc.

Based on the argument above, we decided to adopt the Galactic absorption
for the west lobe.
The excess spectrum is successfully described 
with the PL model modified by Galactic absorption
as is indicated by the dashed lines in the right panel of 
figure \ref{fig:spec_WL} ($\chi^2/{\rm dof} = 142.2/132$).
The best-fit spectral parameters for the excess are tabulated 
in table \ref{tab:spec_WL} (Case 1). 
The photon index and flux density at 1 keV is measured 
as $\Gamma = 2.23_{-0.38-0.12}^{+0.44+0.14}$
and $S_{\rm X} = 15.3_{-3.1}^{+3.0}\pm1.8$ nJy, respectively. 

In figure \ref{fig:SED},
the spectral energy distribution 
of the X-ray excess from the west lobe is compared 
with  that of the radio synchrotron emission.
The radio data of the west lobe are taken from \citet{GRG_image}.
Since the X-ray spectral integration region adopted here 
(the ellipse in figure \ref{fig:img_WL}) corresponds 
to the combination of the {\it western lobe} and {\it western hot spot} regions 
defined in \citet{GRG_image},
we evaluated the radio flux of the west lobe 
by adding those from these two regions. 
The low-frequency synchrotron radio photon index of the west lobe 
was estimated as $\Gamma_{\rm R} = 1.74 \pm 0.07$ 
in the $326$ -- $2695$ MHz range,
where the radio image reveals a dominance of diffuse emission. 
The best-fit index of the X-ray excess from the west lobe 
agrees with this value, within the errors. 
Therefore, we re-examined the PL model 
with the photon index fixed at the radio value ($\Gamma = 1.74$).
As a result, a reasonable fit 
($\chi^2/{\rm dof} = 146.8/133$; Case 2 in table \ref{tab:spec_WL})
was obtained. 
The flux density of the PL component was derived 
as $S_{\rm X} = 12.3 \pm 2.0 \pm 1.9$ nJy at 1 keV. 

\begin{table}[t]
\caption{Summary of the PL fitting to the excess emission from the west lobe}
\label{tab:spec_WL}
\begin{center}
\begin{tabular}{lll}
\hline  
Parameter  &  Case 1   
           & Case 2 \\
\hline  
$N_{\rm H}$ (cm$^{-2}$)
           & \multicolumn{2}{c}{$9.3 \times 10^{19}$\footnotemark[$\dagger$]}\\
$\Gamma $  & $2.23_{-0.38-0.12}^{+0.44+0.14}$\footnotemark[$\S$] 
           & $1.74$\footnotemark[$\ddagger$]\\
$S_{\rm X}$ (nJy)\footnotemark[$*$]  
           & $15.3_{-3.1}^{+3.0}\pm1.8$\footnotemark[$\S$]           
           & $12.3 \pm 2.0 \pm 1.9$\footnotemark[$\S$]  \\
$\chi^2/{\rm dof}$  
           & $142.2 / 132$          
           & $146.8 / 133$\\
\hline        
\multicolumn{3}{@{}l@{}}{\hbox to 0pt{\parbox{71mm}{\footnotesize
\vspace{0.3cm}
\par\noindent\footnotemark[$*$] 
    Flux density at 1 keV  
\par\noindent\footnotemark[$\dagger$] 
    Fixed at the Galactic value \citep{NH}
\par\noindent\footnotemark[$\ddagger$] 
    Fixed at the radio photon index in the $326$ -- $2695$ MHz range. 
\par\noindent\footnotemark[$\S$] 
    The first and second errors represent the statistical and systematic ones, 
    respectively.
}\hss}}\end{tabular}
\end{center}
\end{table}

\section{Discussion} 
\label{sec:discuss}
\subsection{Origin of the X-ray emission from the west lobe} 
\label{sec:origin}
\begin{figure}[h]
\centerline{
\FigureFile(80mm,80mm){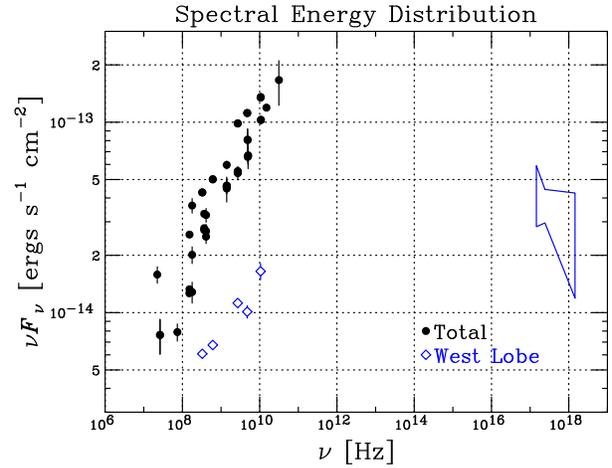}}
\vspace{0.2cm}
\caption{Radio and X-ray spectral energy distribution of 3C 236. 
The PL model best-fitted to the excess X-ray emission 
from the west lobe region, 
corresponding to Case 1 in table \ref{tab:spec_WL},
is shown with the bow tie. 
Only the statistical errors of $\Gamma$ and $S_{\rm X}$ 
are taken into account.
The synchrotron radio data of the west lobe,
estimated by adding the radio flux 
of the "western lobe" and "western hot spot" in \citet{GRG_image}, 
are plotted with the open diamonds.
The filled circles indicate the total radio intensity from 3C 236 
\citep{3C236_radio_1,3C236_radio_2,3C236_radio_3,3C236_radio_4,
3C236_radio_5,3C236_radio_6,3C236_radio_7,3C236_radio_8,
3C236_radio_9,3C236_radio_10,3C236_radio_11,3C236_radio_12,3C236_radio_13,
GRG_image,
3C236_radio_14,3C236_radio_15,3C236_radio_16,
3C236_radio_17,3C236_radio_18,3C236_radio_19}. 
}
\label{fig:SED}
\end{figure}

After carefully subtracting the NXB, XRB and source contamination,
we have detected the excess X-ray emission,
associated with the west lobe of the giant radio galaxy 3C 236,
with a high significance.
Its X-ray spectrum is approximated by the PL model, 
and the photon index was determined 
as $\Gamma = 2.23_{-0.38-0.12}^{+0.44+0.14}$.
This is found to be consistent with 
the $326$ -- $2695$ MHz synchrotron radio index $\Gamma_{\rm R} = 1.74 \pm 0.07$,
when the statistical and systematic errors are considered. 
Thus, a reasonable fit was derived, 
by fixing the photon index at $\Gamma_{\rm R}$.
As shown in figure \ref{fig:SED},
the X-ray spectrum of the west lobe 
does not smoothly connect to the synchrotron radio one. 
Therefore, the X-ray emission is not attributable 
to the highest frequency end of the PL-like synchrotron spectrum 
from the same electron population as for the radio emission.
Referring to the previous X-ray studies on lobes of giant radio galaxies 
(e.g., \cite{3C326,3C35,DA240}),
these properties are thought to support an IC origin for this X-ray emission. 

\begin{longtable}[t]{llll}
\caption{Summary of Energetics in the west lobe of 3C 236}
\label{tab:spec_UeUm}
\hline  
\multicolumn{2}{l}{Parameter} &  Value                  & Comments \\
\hline        
\endfirsthead
\hline  
\multicolumn{2}{l}{Parameter} &  Value                  & Comments \\
\hline        
\endhead
\endfoot
\endlastfoot
Input &$S_{\rm X}$ (nJy)        & $12.3 \pm 2.0 \pm 1.9$ & Case 2\\
      &$S_{\rm R}$ (Jy)         & $1.11 \pm 0.02$        & at $608.5$ MHz\\
      &$\Gamma_{\rm R}$         & $1.74 \pm 0.07$        & in $326$ -- $2695$ MHz\\
      &$V$ (cm$^3$)           & $8.5 \times 10^{72}$     & \\
\hline        
Output &$u_{\rm e}$ ($10^{-14}$ erg cm$^{-3}$) 
                              & $3.9_{-0.7 -0.9}^{+0.6 +1.0}$  &  $\gamma_{\rm e} = 10^{3}$ -- $10^{5}$\\
       &                      & $12.6_{-2.1 -2.1}^{+2.0 +2.3}$  &  $\gamma_{\rm e} = 10^{2}$ -- $10^{5}$\\ 
       &$u_{\rm m}$ ($10^{-14}$ erg cm$^{-3}$) 
                              & $0.92_{-0.15 -0.35}^{+0.21 +0.52}$ & \\
       &$B$ ($\mu$G)          & $0.48_{-0.04 -0.10}^{+0.05 +0.12}$ & \\
       &$u_{\rm e}/u_{\rm m}$    & $4.2_{-1.3 -2.3}^{+1.6 +4.1}$   & $\gamma_{\rm e} = 10^{3}$ -- $10^{5}$\\
       &                      & $13.7_{-4.3 -5.7}^{+5.3 +8.3}$ & $\gamma_{\rm e} = 10^{2}$ -- $10^{5}$\\
\hline        
\end{longtable}

The synchrotron X-ray hypothesis is further rejected 
by an argument on the electron radiative time scale.
We evaluate the synchrotron life time of electrons 
which could produce the synchrotron X-ray photons,
by simply assuming the minimum energy condition \citep{Bme}.
Because the magnetic field strength under this assumption 
is found to be not so far from the more realistic value 
to be discussed in \S\ref{sec:energetics}, 
the following discussion basically holds.
Corresponding to the emission region defined in \S \ref{sec:image},
the three-dimensional shape of the west lobe is assumed to be 
an ellipsoid with a major and minor radius of $740$ kpc and $305$ kpc, 
respectively. 
From the synchrotron radio flux density, 
$S_{\rm R} = 1.11$ Jy at 608.5 MHz (calculated from \cite{GRG_image}),
and the spectral slope of $\Gamma_{\rm R} = 1.74$, 
the minimum energy magnetic field in the west lobe 
was estimated as $B_{\rm me} \sim 0.8 $ $\mu$G,
neglecting the proton contribution.
A similar magnetic field strength was derived by \citet{GRG_WENSS} 
under the equipartition condition as $B_{\rm eq} = 0.86 $ $\mu$G.
In this magnetic field, 
the synchrotron photons at 1 keV are expected to be radiated from 
electrons with a Lorentz factor of $\gamma_{\rm e} \sim 5 \times 10^{8}$. 
Such electron is estimated to loose half of their energy in a time scale of 
$T_{\rm sync} = \frac{3 m_{\rm e} c}{4 \sigma_{\rm T} u_{\rm me}} \gamma_{\rm e}^{-1}
            \sim 0.08$ Myr
through the synchrotron radiation, 
where $m_{\rm e}$ is the electron mass, 
$c$ is the speed of light, $\sigma_{\rm T}$ is the Thomson cross section,
and
$u_{\rm me} = B_{\rm me}^2/8\pi$ is the minimum-energy magnetic energy density.
It is important to note that 
this $T_{\rm sync}$ value overestimates the actual radiative cooling time scale
in the lobe of 3C 236, since the Compton loss is neglected.
Nevertheless,
the time scale is by three orders of magnitude shorter 
than the spectral age of 3C 236, 
$T_{\rm age} = 98\pm3$ Myr \citep{GRG_WENSS}.
We, thus, regard the synchrotron X-ray explanation 
as encountering a serious problem due to the significant radiative loss.

Based on the above consideration,
we safely ascribe the detected X-ray emission to the IC radiation 
from the synchrotron-emitting electrons within the west lobe of 3C 236.
This result has made 3C 236 the largest radio galaxy 
of which the lobe is ever studied through the IC photons. 

In the case of radio lobes, 
three candidate seed photon sources for the IC scattering 
were widely proposed; the synchrotron radio photons in the lobes themselves,
the infra-red radiation field from the nucleus \citep{IC_nuclearIR},
and the CMB radiation \citep{CMB_IC}.
The energy density of the synchrotron radiation 
spatially averaged over the west lobe 
is estimated to be very low as $\gtrsim 10^{-19}$ erg cm$^{-3}$ 
from the radio flux density $S_{\rm R} = 1.11$ Jy at 608.5 MHz.
The infra-red flux density from the nucleus of 3C 236,
$8.4$ mJy at the K-band \citep{2MASS},
is converted to the nuclear monochromatic K-band luminosity 
as $ 3 \times 10^{44}$ ergs s$^{-1}$.
This nuclear photon field yields an infrared energy density 
of $4 \times 10^{-16}$ ergs s$^{-1}$ at the nearest edge of the west lobe,
which is located at the projected distance of $\sim 450$ kpc from the nucleus. 
At the redshift of 3C 236 ($z = 0.100500$; \cite{3C236_redshift}),
the CMB energy density is precisely predicted 
as $u_{\rm CMB} = 6.0\times 10^{-13}$ ergs s$^{-1}$. 
Therefore, we have concluded that 
the seed photons are dominantly provided by the CMB radiation
and the other candidate sources have only a negligible contribution.

\subsection{Energetics in the west lobe of 3C 236} 
\label{sec:energetics}
By comparing the IC X-ray and synchrotron radio spectral energy distributions
presented in figure \ref{fig:SED},
we can measure the energy densities of the electrons and magnetic field,
$u_{\rm e}$ and $u_{\rm m}$ respectively, in the west lobe of 3C 236.
The input observables, required to calculate $u_{\rm e}$ and $u_{\rm m}$,
are listed in the upper columns of table \ref{tab:spec_UeUm}. 
The synchrotron radio flux density and photon index 
($S_{\rm R} = 1.11\pm 0.02$ Jy at 608.5 MHz and 
$\Gamma_{\rm R} = 1.74 \pm 0.07$ in the $326$ -- $2695$ MHz range, respectively) 
were derived from \citet{GRG_image}.
We adopted the IC X-ray flux density at 1 keV of 
$S_{\rm X} = 12.3 \pm 2.0 \pm 1.9$ nJy
derived in Case 2 where the X-ray photon index was fixed at the radio value. 
Corresponding to the radio and X-ray spectral slope,  
the electron number density spectrum was assumed to be 
a simple PL form described as $\propto \gamma_{\rm e}^{-(2 \Gamma_{\rm R} -1) }$. 
The volume of the west lobe was estimated as $V = 8.5 \times 10^{72}$ cm$^{3}$,
since we approximated the shape of the west lobe 
to be the ellipsoid with the major and minor radius of 
$740$ kpc and $305$ kpc, respectively 
(see \S \ref{sec:image} and \S \ref{sec:origin}).

We analytically evaluated the energetics in the west lobe of 3C 236
by referring to \citet{CMB_IC}.
The result is summarized in the lower columns of table \ref{tab:spec_UeUm}. 
Here and hereafter, 
the statistical error of $S_{\rm X}$ is propagated to the first one,
while all the other errors
(including those in $S_{\rm R}$, $\Gamma_{\rm R}$ and so forth)
are taken into account in the second one. 
The specially averaged magnetic field strength 
is derived as $B = 0.48_{-0.04 -0.10}^{+0.05 +0.12}$ $\mu$G
from the equation (11) in \citet{CMB_IC} described as 
\begin{eqnarray}
B^{\alpha_{\rm R}+1} = 
  \frac{(5.05\times10^4)^{\alpha_{\rm R}} C(\alpha_{\rm R}) G(\alpha_{\rm R}) 
        (1+z)^{\alpha_{\rm R}+3}} {10^{47}}
       \frac{S_{\rm R}}{S_{\rm X}}
       \left( \frac{\nu_{\rm R}}{\nu_{\rm X}}\right)^{\alpha_{\rm R}} 
       \nonumber
\end{eqnarray},
where $\alpha_{\rm R} = \Gamma_{\rm R} - 1$ is the synchrotron radio 
(and hence the IC X-ray) energy index, 
$\nu_{\rm R}$ and $\nu_{\rm X}$ are the frequencies 
at $S_{\rm R}$ and $S_{\rm X}$ are measured respectively,
$C(\alpha_{\rm R})$ is approximately constant at $\sim 1.15 \times 10^{31}$
over the range of $\alpha_{\rm R} = 0.5$ -- $2.0$, 
and $G(\alpha_{\rm R})$ is evaluated as $\simeq 0.5$ 
at $\alpha_{\rm R} = 0.74$ ($\Gamma_{\rm R} = 1.74$).
This magnetic field strength is equivalent to the magnetic energy density of 
$u_{\rm m} = 0.92_{-0.15 -0.35}^{+0.21 +0.52}\times 10^{-14} $ ergs cm$^{-3}$.

For the electron spectral slope of $(2 \Gamma_{\rm R} -1)> 2$ 
(i.e., $\Gamma_{\rm R} > 1.5$), 
the $u_{\rm e}$ estimate is known to be rather sensitive 
to the assumption on the minimum Lorentz factor of the electrons.
However, it has not yet been observationally explored. 
Following the previous studies (e.g., \cite{DA240}),
we first evaluated the energy density of electrons 
with a Lorentz factor of $\gamma_{\rm e} = 10^{3}$ -- $10^{5}$,
because they are directly observable through the IC and synchrotron radiations.
The energy density of these electrons was derived as 
$u_{\rm e} = 3.9_{-0.7 -0.9}^{+0.6 +1.0} \times 10^{-14} $ ergs cm$^{-3}$.
This indicates a slight electron dominant condition in the west lobe,
as parameterized by the electron to magnetic field energy density radio 
of $u_{\rm e}/u_{\rm m} = 4.2_{-1.3 -2.3}^{+1.6 +4.1}$.
Second, 
we lowered the minimum electron Lorentz factor down to $\gamma_{\rm e} = 10^{2}$,
since the existence of such low-energy electrons 
was suggested in a number of strong and compact radio galaxies 
\citep{IC_nuclearIR}. 
In addition, a similar value of the minimum Lorentz factor was adopted 
to interpret low-frequency radio data of some giant radio galaxies 
\citep{LowFreq_GRG}.
As a result, the electron energy density increased to 
$u_{\rm e} = 12.6_{-2.1 -2.1}^{+2.0 +2.3} \times 10^{-14}$ erg cm$^{-3}$.
Correspondingly, the electron dominance in the lobe 
was found to become more prominent,
as $u_{\rm e}/u_{\rm m} = 13.7_{-4.3 -5.7}^{+5.3 +8.3}$.

\subsection{Overall picture} 
\label{sec:evolution}
\begin{figure}[h]
\centerline{
\FigureFile(80mm,80mm){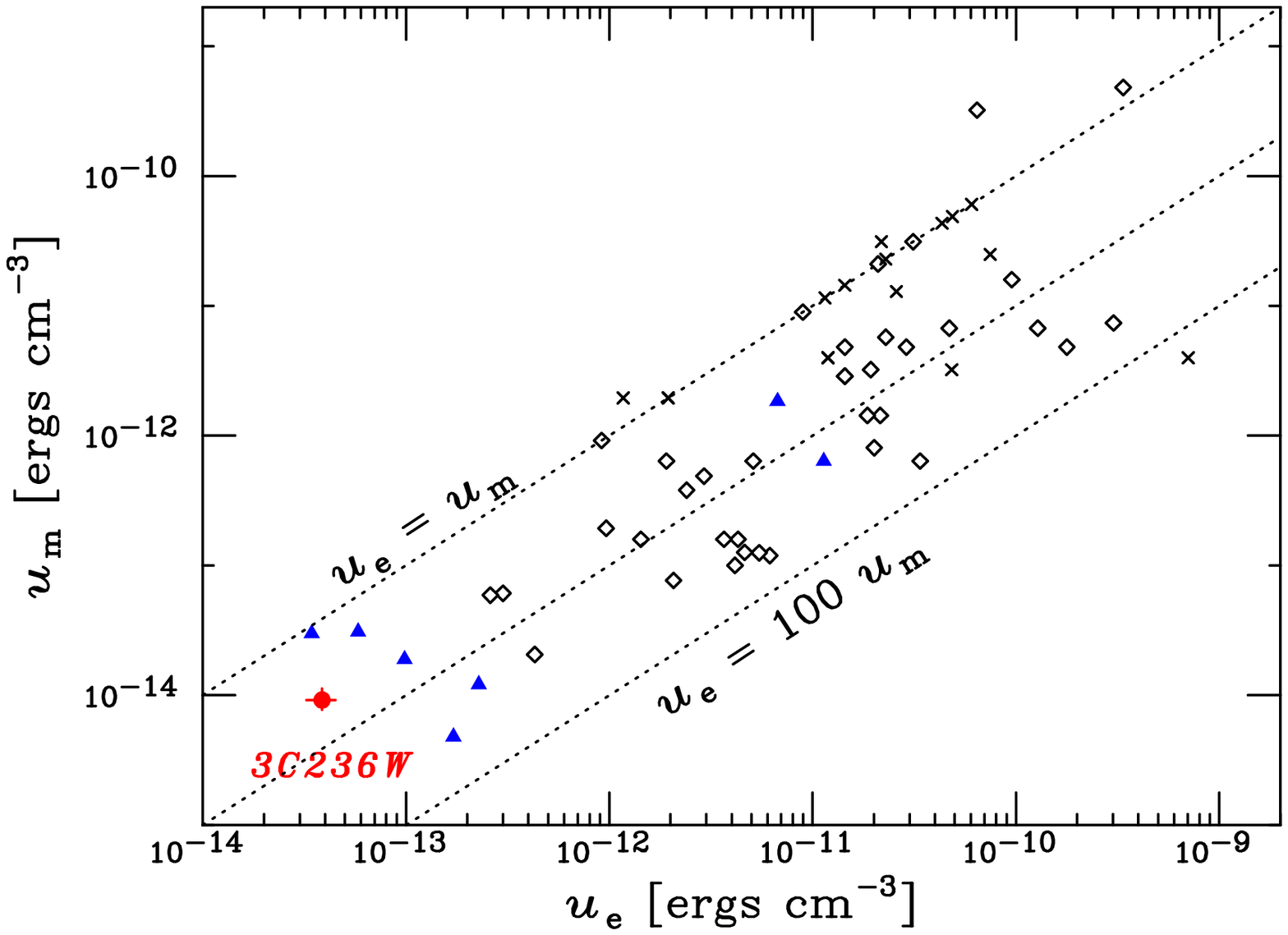}}
\vspace{0.2cm}
\caption{
$u_{\rm e}$ -- $u_{\rm m}$ relation in lobes of radio galaxies
(\cite{lobes_Croston,3C326,3C35,DA240,3C457_XMM}, and reference therein).
$u_{\rm e}$ is evaluated 
for the electron Lorentz factor of $\gamma_{\rm e} = 10^{3}$ -- $10^{5}$.
The 3C 236 west lobe is pointed by the filled red circle.
The lobes of giant radio galaxies ($D\gtrsim 1$ Mpc) 
are plotted with the filled blue triangles, 
while those of moderate radio galaxies are indicated by the open diamonds.
For lobes shown with the crosses, 
only the upper limit on the IC X-ray flux 
(correspondingly the upper and lower limits 
on $u_{\rm e}$ and $u_{\rm m}$, respectively) was determined.
The diagonal dotted lines represent the condition of 
$u_{\rm e} / u_{\rm m} = 1 $, $10$, and $100$.
}
\label{fig:ue-um}
\end{figure}

Figure \ref{fig:ue-um} compiles the relation 
between $u_{\rm e}$ (for $\gamma_{\rm e} = 10^{3}$ -- $10^{5}$) and $u_{\rm m}$
in the lobes of radio galaxies, 
from which the IC X-ray emission was detected 
(\cite{lobes_Croston,3C457_XMM,DA240}, and reference therein).
The figure clearly demonstrates 
that a typical electron dominance of $u_{\rm e}/u_{\rm m} = 1$ -- $100$ 
is realized in the radio lobes,
regardless of their physical size \citep{3C452,3C98,ForA,3C326,3C35,DA240}. 
Our Suzaku result on the west lobe of the giant radio galaxy 3C 236
is found to be fully compatible with this picture. 

Due to their low energy densities, 
the IC-detected giant radio galaxies are distributed 
in the bottom-left end of the $u_{\rm e}$--$u_{\rm m}$ plot 
(i.e., figure \ref{fig:ue-um}).
In the discussion, 
we neglect the lobes of the giant radio galaxy, 3C 457,
which is located around the center in figure \ref{fig:ue-um},
since it is reported to be one of the youngest sources 
($T_{\rm age } \sim 30$ Myr; \cite{3C457_XMM}) among the giant radio galaxies.
In the lobes of giant radio galaxies,
a dominance of the IC radiative loss over the synchrotron one is 
theoretically pointed out,
because their magnetic energy density is predicted to be less 
than the CMB energy density  \citep{GRG_ICdominance}. 
The recent Suzaku results on the giant radio lobes \citep{3C326,3C35,DA240},
including our present result on 3C 236, 
have observationally confirmed this idea, 
by demonstrating a condition of $u_{\rm m}/ u_{\rm CMB} < 0.1$. 
Especially, the IC loss is extremely dominant in 3C 236 
as $u_{\rm m}/u_{\rm CMB} \sim 0.01$. 

\begin{figure}[t]
\centerline{
\FigureFile(80mm,80mm){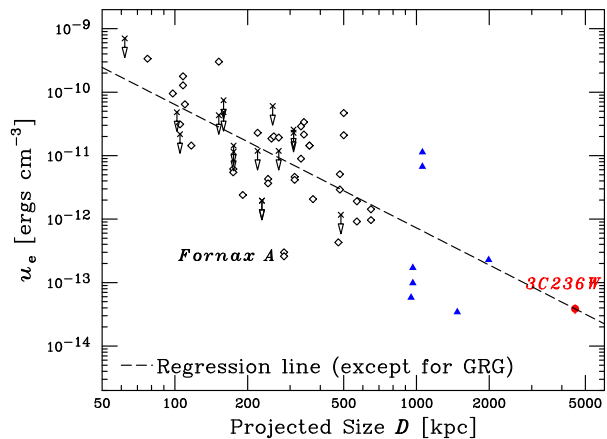}}
\vspace{0.2cm}
\caption{
Electron energy density in the lobes, $u_{\rm e}$,
plotted as a function of the total projected length, $D$, 
of the radio galaxies.  
The same symbol/color notation as for figure \ref{fig:ue-um} is adopted.
The dashed line indicate the regression line, $u_{\rm e} \propto D^{-1.9}$,
for radio galaxies with a moderate size ($D\lesssim 1$ Mpc). 
}
\label{fig:ue-length}
\end{figure}

In figure \ref{fig:ue-length}, 
the electron energy density in the lobes are plotted 
against the total projected length of the radio galaxies, $D$.
\citet{3C326} proposed that a lobe evolves 
along a track of $u_{\rm e} \propto D^{-2}$ on the plot,
while it is actively energized by its jet.
The lobe is also predicted to deviate from the $u_{\rm e}$--$D$ track
toward a lower $u_{\rm e}$ value, 
due to significant radiative and adiabatic losses, 
after the energy import by its jet has ceased. 
The relation of $u_{\rm e} \propto D^{-1.9\pm 0.4}$ observed 
for the radio galaxies with a moderate size of $ D \lesssim 900 $ kpc 
(the dashed line in figure \ref{fig:ue-length}; \cite{DA240})
is thought to qualitatively back up the former idea.
The latter scenario was observationally demonstrated 
by the radio galaxy Fornax A
which is reported to host a very dormant nucleus and jet. 
As is shown in figure \ref{fig:ue-length},
the electron energy density in the lobes of Fornax A 
is by more than a factor of 10 smaller than the value 
expected from the $u_{\rm e}$--$D$ regression line for 
the radio galaxies with $ D < 900 $ kpc.

The recent systematic studies with Suzaku (e.g., \cite{3C35,DA240})
revealed that the lobes of the giant radio galaxies tend to exhibit 
a lower $u_{\rm e}$ value by nearly an order of magnitude, 
in comparison to the simple extension of the regression line 
for the $D \lesssim 900$ kpc radio galaxies.
This is clearly visualized in figure \ref{fig:ue-length}.
By analogy with Fornax A, this property indicates that 
the activity of the jets in the giant radio galaxies 
have been already declined, 
and their energy input to the lobes is currently defeated by 
the radiative and adiabatic losses of the electrons \citep{3C35,DA240}.
In contrast, $u_{\rm e}$ in the west lobe of 3C 236 coincides with 
the value estimated from the $u_{\rm e}$--$D$ relation for the smaller sources.
This result suggests that 
the lobe is still supplied with a sufficient energy from the jet.
Even though the current lobe is energetically decoupled from the jet, 
it has not yet entered a cooling-dominant regime.

The ongoing energy transport (or transport until recently)
to the west lobe of 3C 236
appears to be reinforced by the observational fact 
that the radio spectrum around the west hot spot
(e.g., $\Gamma_{\rm R} = 1.69$ in $326$ MHz -- $10.6$ GHz 
for the {\itshape western hot spot} region in \cite{radio_spec_map})
is harder than those in the surrounding regions.
\citet{3C236_radio_structure} closely investigated similarities 
between the small-scale ($\sim 2$ kpc) and 
large-scale ($\sim$ Mpc) radio structures.
They pointed out that the interaction of the jet with the interstellar medium 
in the central region of 3C 236, creating the small-scale structure, 
continued for more than an order of magnitude longer than $\sim 10$ Myr.
The interaction is inferred to have consequently formed
the the Mpc-scale lobes.
The energy supply to the lobe is possible 
to be maintained by such a long-lasting inteaction.

In contrast, the lack or weekness of the radio jet 
connecting the radio structures in small and large scales
especially toward the west direction \citep{3C236_radio_structure},
which is reminiscent of giant double-double radio galaxies
\citep{B1144+352_double-double}, 
implies that the jet activity in 3C 236 is highly variable or episodic,
rather than stable over its life time
(e.g., \cite{3C236_interrupted_activity,3C236_nuclear_structure}). 
By comparing the dynamical age of the small-scale structure ($\gtrsim 0.1$ Myr)
and the radiative age of young electrons contained 
in the hot spots within the large-scale lobes \citep{radio_spec_map},
\citet{3C236_interrupted_activity} proposed
that the jet had been possibly switched off for $\sim 10$ Myr.
A high-resolution VLBI image \citep{3C236_nuclear_structure} 
supports a recent ignition of a one-sided jet to the west 
in a 10 mas scale (corresponding to $\sim 20$ pc at the 3C 236 rest frame).
The jet reactivation was suggested to be triggered $\sim 0.1$ Myr ago 
by a minor merger event that took place $\sim 10$ Myr ago 
\citep{3C236_AGN_feedback}. 

We briefly examine the impact of adiavatic and radiative losses 
during the suggested jet hibernation.
When the distance from the nucleus to the lobe edge ($\sim 1.6$ Mpc)
is devided by the source age ($T_{\rm age} = 98$ Myr; \cite{GRG_WENSS}), 
the time-averaged advance speed of the 3C 236 west lobe is roughly evaluated 
as $\sim 0.05c$. 
This agrees with the typical lobe speed 
measured for Fanaroff-Riley II radio galaxies 
($0.03c$ -- $0.1 c$; \cite{lobe_speed}).
At the advance speed, 
the lobe head proceeds to $\sim 160$ kpc in $10$ Myr.
This results in an expansion of the west lobe volume 
by less than $\sim 40$ \%.
Therefore, the adiabatic loss during the jet dormancy 
is considered to be relatively insignificant 
in the 3C 236 west lobe. 
 The radiative time scale of the electrons in the west lobe 
is evaluated as $\sim 16$ Myr $(\gamma_{\rm e}/10^5)^{-1}$ 
where both the synchrotron and IC radiation is taken into account.
Here, the electron Lorents factor of $\gamma_{\rm e} = 10^5$ 
corresponds to the synchrotron photons of $\sim 5$ GHz
under the magnetic field strength of $B=0.48$ $\mu$G
which is measured in this study.
This means that the radiative cooling due to the jet hibernation
is neglidgible to the electrons,
for which we integrated the energy density $u_{\rm e}$ 
(i.e., $\gamma_{e} = 10^3$ -- $10^5$ or $10^2$ -- $10^5$).
As mentioned above, 
the flat synchrotron spectrum in the hot spot 
is compatible with the scenario that 
the west lobe of 3C 236 has not yet been in the cooling domain.
Therefore, we have concluded that our result does not 
contradict to the reported suspention of the jet ejection.

Another energy injection channel is hinted in the 1.4 GHz radio image 
by \citet{3C236_radio_structure},
which reveals a wiggling inner ridge in the central region of the west lobe. 
They interpreted that the rigde was created 
by a recent jet penetrating the west lobe. 
This possibly yields a very attractive idea 
that the new jet re-energizes the west lobe 
and prevents it from being cooled in the future. 

\vspace{0.5cm}
We thank the anonymous reviewer 
for her/his constructive and supportive suggestions.
We are grateful to all the members of the Suzaku team,
for the successful operation and calibration.
This research has made use of the archival Chandra data and
its related software provided by the Chandra X-ray Center (CXC).


\clearpage

\clearpage


\begin{thebibliography}{}
\bibitem[Abdo et al.(2012)]{CenA_Fermi}
  Abdo, A. A. et al., 2010, Science, 328, 725
\bibitem[Adgie et al.(1972)]{3C236_radio_1}
  Adgie, R. L., Crowther, J. H., \& Gent, H.
  1972, \mnras, 159, 233
\bibitem[Alexander \& Leahy(1987)]{lobe_speed}
  Alexander, P., \& Leahy, J. P., 1987, \mnras, 225, 1
\bibitem[Barthel et al.(1985)]{3C236_radio_structure}
  Barthel, P. D., Miley, G. K., Jagers, W. J., Schilizzi, R. T., 
  \& Strom, R. G., 1985, \aap, 148, 243
\bibitem[Becker et al.(1991)]{3C236_radio_2}
  Becker, R. H., White, R. L., \& Edwards, A. L.,
  1991, \apjs, 75, 1
\bibitem[Becker et al.(1995)]{3C236_radio_3}
  Becker, R. H.; White, R. L., \& Helfand, D. J.,
  1995, \apj, 450, 559
\bibitem[Brunetti et al.(1997)]{IC_nuclearIR}
  Brunetti, G., Setti, G., \& Comastri, A.,
  1997, \aap, 325, 898
\bibitem[Cohen et al.(2007)]{3C236_radio_4}
  Cohen, A. S., et al. 
  2007, \aj, 134, 1245
\bibitem[Colla et al.(1973)]{3C236_radio_5}
  Colla, G., et al.,
  1973, \aaps, 11, 291
\bibitem[Condon et al.(1998)]{3C236_radio_6}
  Condon, J. J., et al., 
  1998, \aj, 115, 1693
\bibitem[Croston et al.(2005)]{lobes_Croston}
  Croston, J. H., Hardcastle, M. J., Harris, D. E., 
  Belsole, E., Birkinshaw, M., \& Worrall, D. M.,
  2005, \apj, 626, 733
\bibitem[Douglas et al.(1996)]{3C236_radio_7}
  Douglas, J. N., Bash, F. N., Bozyan, F. A., Torrence, G. W., \& Wolfe, C.,
  1996, \aj, 111, 1945
\bibitem[Feigelson et al.(1995)]{ForA_ROSAT}
  Feigelson, E. D., Laurent-Muehleisen, S. A.,
  Kollgaard, R. I., \& Fomalont, E. B.,
  1995, \apj, 449, L149 
\bibitem[Genzel et al.(1976)]{3C236_radio_8}
  Genzel, R., Pauliny-Toth, I. I. K., Preuss, E., \& Witzel, A.,
  1976, \aj, 81, 1084
\bibitem[Gregory \& Condon(1991)]{3C236_radio_9}
  Gregory, P. C., \& Condon, J. J.,
  1991, \apjs, 75, 1011
\bibitem[Hales et al.(1988)]{3C236_radio_10}
  Hales, S. E. G., Baldwin, J. E., \& Warner, P. J.,
  1988, \mnras, 234, 919
\bibitem[Harris \& Grindlay(1979)]{CMB_IC}
  Harris, D. E., \&  Grindlay, J. E., 1979, \mnras, 188, 25
\bibitem[Hill et al.(1996)]{3C236_redshift}
  Hill, G. J., Goodrich, R. W., \& Depoy, D. L.,
  1996, \apj, 462, 163
\bibitem[Ishisaki et al.(2007)]{xissimarfgen}
  Ishisaki, Y.,  et al. 2007, \pasj, 59, 113
\bibitem[Ishwara-Chandra \& Saikia(1999)]{GRG_ICdominance}
  Ishwara-Chandra, C. H., \& Saikia, D. J.,
  1999, \mnras, 309, 100
\bibitem[Isobe et al.(2002)]{3C452}
  Isobe, N., et al., 2002, \apj, 580, L111
\bibitem[Isobe et al.(2009)]{3C326}
  Isobe, N., et al., 2009, \apj, 706, 454
\bibitem[Isobe et al.(2005)]{3C98}
  Isobe, N., Makishima, K., Tashiro, M., \& Hong, S., 2005, \apj, 632, 781
\bibitem[Isobe et al.(2006)]{ForA}
  Isobe, N., Makishima, K., Tashiro, M., Itoh, K.,
  Iyomoto, N., Takahashi, I., \& Kaneda, H., 2006, \apj, 645, 256
\bibitem[Isobe et al.(2011a)]{3C35}
  Isobe, N., Seta, H., Gandhi, P., \& Tashiro, M.S., 
  2011a, \apj, 727, 82
\bibitem[Isobe et al.(2011b)]{DA240}
  Isobe, N., Seta, H., \& Tashiro, M.S.,
  2011b \pasj, 63, S947
\bibitem[Kalberla et al.(2005)]{NH}
  Kalberla, P. M. W., Burton, W. B., Hartmann, Dap, 
  Arnal, E. M., Bajaja, E., Morras, R., \& Po\"{o}ppel, W. G. L.,
  2005,\aap, 440, 775
\bibitem[Kaneda et al.(1995)]{ForA_ASCA}
  Kaneda, H., et al., 1995, \apj, 453, L13  
\bibitem[Kellermann et al.(1969)]{3C236_radio_11}
  Kellermann, K. I., Pauliny-Toth, I. I. K. \& Williams, P. J. S.,
  1969, \apj, 157, 1
\bibitem[Kellermann \& Pauliny-Toth(1973)]{3C236_radio_12}
  Kellermann, K. I. \& Pauliny-Toth, I. I. K.,
  1973, \aj, 78, 828K
\bibitem[Konar et al.(2010)]{3C457_XMM}
  Konar, C., Hardcastle, M. J., Croston, J. H., \& Saikia, D. J.
  2009, \mnras, 400, 480
\bibitem[Koyama et al.(2007)]{XIS}
  Koyama, K., et al., 2007, \pasj, 59, S23
\bibitem[Kuhr et al.(1981)]{3C236_radio_13}
  K\"uhr, H., Witzel, A.,  Pauliny-Toth, I. I. K., \& Nauber, U.,
  1981, \aaps, 45, 367
\bibitem[Kushino et al.(2002)]{XRB_ASCA}
  Kushino, A., Ishisaki, Y.,  Morita, U., Yamasaki, N. Y.,
  Ishida, M., Ohashi, T., \& Ueda, Y., 
  2002, \pasj, 54, 327
\bibitem[Laing et al.(1983)]{RG_178MHz}
  Laing, R. A., Riley, J. M., \& Longair, M. S.,
  1983, \mnras, 204, 151
\bibitem[Labiano et al.(2013)]{3C236_AGN_feedback}
  Labiano, A., et al., 2013, \aap, 549, A58
\bibitem[Lumb et al.(2002)]{XRB_XMM}
  Lumb, D. H., Warwick, R. S., Page, M., \& De Luca, A., 
  2002, \aap, 389, 93  
\bibitem[Machalski et al.(2008)]{J1420}
  Machalski, J., Koziel-Wierzbowska, D., Jamrozy, M., \& Saikia, D. J.,
  2008, \apj, 679, 149
\bibitem[Mack et al.(1997)]{GRG_image}
  Mack, K.-H., Klein, U., O'Dea, C. P., \& Willis, A. G.,
  1997, \aaps, 123, 423,
\bibitem[Mack et al.(1998)]{radio_spec_map}
  Mack et al., 1998, A\&A, 329, 431
\bibitem[Mewe et al.(1985)]{MEKAL}
  Mewe, R., Gronenschild, E. H. B. M., \& van den Oord, G. H. J.
  1985, \aaps, 62, 197
\bibitem[Migliori et al.(2007)]{PicA_XMM}
  Migliori, G., Grandi, P., Palumbo, G. G. C.,
  Brunetti, G. \& Stanghellini, C.
  2007, \apj, 668, 203
\bibitem[Miley(1980)]{Bme}
  Miley, G., 1980, \araa, 18, 165 
\bibitem[Mitsuda et al.(2007)]{Suzaku}
  Mitsuda, K., et al., 2007, \pasj,  59, S1 
\bibitem[Miyazawa et al.(2009)]{AGN_Suzaku}
  Miyazawa, T., Haba, Y., \& Kunieda, H.
  2009, \pasj, 61, 1331
\bibitem[Morganti et al.(2005)]{3C236_outflow}	
  Morganti, R., Tadhunter, C. N., \& Oosterloo, T. A.
  2005, \aap, 444, L9
\bibitem[O'Dea et al.(2001)]{3C236_interrupted_activity}
  O'Dea, C. P., et al., 2001, \aj, 121, 1915
\bibitem[Orr\`{u} et al.(2010)]{LowFreq_GRG}
  Orr\`{u}, E., et al.
  2010, \aap, 515, 50
\bibitem[Pauliny-Toth et al.(1966)]{3C236_radio_14}
  Pauliny-Toth, I. I. K.,  Wade, C. M., \& Heeschen, D. S.,
  1966, \apjs, 13, 65
\bibitem[Pilkington et al.(1965)]{3C236_radio_15}
  Pilkington, J. D. H. \& Scott, J. F.,
  1965, \memras, 69, 183P
\bibitem[Roger et al.(1986)]{3C236_radio_16}
  Roger, R. S., Costain, C. H., \& Stewart, D. I.,
  1986, \aaps, 65, 485
\bibitem[Schilizzi et al.(2001)]{3C236_nuclear_structure}
  Schilizzi, R. T., et al., 2001, \aap, 368, 398
\bibitem[Schoenmakers et al.(1999)]{B1144+352_double-double}
  Schoenmakers, A. P., de Bruyn, A. G., 
  R\"{o}ttgering, H. J. A., \& van der Laan, H.,
  1999, \aap, 341, 44
\bibitem[Schoenmakers et al.(2000)]{GRG_WENSS}
  Schoenmakers, A. P., et al., 2000, \aaps, 146, 293
\bibitem[Serlemitsos et al.(2007)]{XRT}
  Serlemitsos, P. J., et al., 2007, \pasj, 59, S9
\bibitem[Skrutskie et al.(2006)]{2MASS}
  Skrutskie, M. F., et al., 2006, \aj, 131, 1163
\bibitem[Stawarz et al.(2013)]{CenA}
  Stawarz, \L., et al., 2013. \apj, 766, 48
\bibitem[Strom \& Willis(1980)]{3C236_largest2}
  Strom, R. G., \& Willis, A. G.,
  1980, \aap, 85, 36
\bibitem[Takahashi et al.(2007)]{HXD}
  Takahashi, T., et al., 2007, \pasj, 59, S35
\bibitem[Takeuchi et al.(2012)]{NGC6251}
  Takeuchi, Y. et al., 2012, \apj, 749, 66
\bibitem[Tashiro et al.(1998)]{CenB}
  Tashiro, M., et al., 1998, \apj, 499, 713
\bibitem[Tashiro et al.(2009)]{ForA_Suzaku}
  Tashiro M., Isobe, N., Seta H., Yaji, Y., \& Matsuta K., 
  2009, \pasj, 61, S327
\bibitem[Tawa et al.(2008)]{xisnxbgen}
  Tawa, N., et al., 2008, \pasj, 60, S11
\bibitem[Viner \& Erickson(1975)]{3C236_radio_17}
  Viner, M. R., \& Erickson, W. C.,
  1975, \aj, 80, 931
\bibitem[Voges et al.(1999)]{1RXS}
  Voges, W., et al.,  1999, \aap, 349, 389
\bibitem[Waldram et al.(1996)]{3C236_radio_18}
  Waldram, E. M.; Yates, J. A.; Riley, J. M.; Warner, P. J.,
  1996, \mnras , 282, 779
\bibitem[Willis et al.(1974)]{3C236_largest}	
  Willis, A. G., Strom, R. G., \& Wilson, A. S.,
  1974, \nat, 250, 625
\bibitem[Witzel et al.(1978)]{3C236_radio_19}
  Witzel, A., Pauliny-Toth, I. I. K., Geldzahler, B. J., \& Kellermann, K. I.
  1978, \aj, 83, 475
\end{thebibliography}
\end{document}